\documentclass[aps,preprint,showpacs]{revtex4}

\usepackage{amsmath,amssymb}

\renewcommand\d{\partial}
\newcommand\+{\dagger}
\newcommand\<{\langle}
\renewcommand\>{\rangle}
\newcommand\x{{\textbf{x}}}
\newcommand\y{{\textbf{y}}}

\begin{document}

\preprint{INT PUB 08-08}

\title{Toward an AdS/cold atoms correspondence: a geometric realization
of the Schr\"odinger symmetry}

\author{D.~T.~Son}
\affiliation{Institute for Nuclear Theory, University of Washington, 
Seattle, Washington 98195-1550, USA}

\date{April 2008}

\begin{abstract}
We discuss a realization of the nonrelativistic conformal group (the
Schr\"odinger group) as the symmetry of a spacetime.  We write down a
toy model in which this geometry is a solution to field equations.  We
discuss various issues related to nonrelativistic holography.  In
particular, we argue that free fermions and fermions at unitarity
correspond to the same bulk theory with different choices for the
near-boundary asymptotics corresponding to the source and the
expectation value of one operator.  We describe an extended version of
nonrelativistic general coordinate invariance which is realized
holographically.
\end{abstract}
\pacs{11.25.Tq, 03.75.Ss}

\maketitle

\section{Introduction}

The anti--de Sitter/conformal field theory (AdS/CFT)
correspondence~\cite{Maldacena:1997re,Gubser:1998bc,Witten:1998qj}
establishes the equivalence between a conformal field theory in flat
space and a string theory in a higher-dimensional curved space.  The
best known example is the equivalence between ${\cal N}=4$
supersymmetric Yang-Mills theory and type IIB string theory in
AdS$_5\times$S$^5$ space.  The strong coupling limit of the field
theory corresponds to the supergravity limit in which the string
theory can be solved.  In the recent literature, the ${\cal N}=4$
supersymmetric Yang-Mills theory at infinite 't Hooft coupling is
frequently used as a prototype to illustrate features of strongly coupled
gauge theories.

There exist, in nonrelativistic physics, another prototype of strong
coupling: fermions at unitarity~\cite{Eagles,Leggett,Nozieres}.  This
is the system of fermions interacting through a short-ranged potential
which is fine-tuned to support a zero-energy bound state.  The system
is scale invariant in the limit of zero-range potential.  Since its
experimental realizations using trapped cold atoms at the Feshbach
resonance~\cite{OHara,Jin,Grimm,Ketterle,Thomas,Salomon}, this system
has attracted enormous interest.

One may wonder if there exists a gravity dual of fermions at
unitarity.  If such a gravity dual exists, it would extend the notion
of holography to nonrelativistic physics, and could potentially bring
new intuition to this important strongly coupled system.  Similarities
between the ${\cal N}=4$ super--Yang-Mills theory and unitarity
fermions indeed exist, the most important of which is scale invariance.
The have been some speculations on the possible relevance of the
universal AdS/CFT value of the viscosity/entropy density
ratio~\cite{Kovtun:2004de} for unitarity
fermions~\cite{Gelman:2004fj,Schafer:2007pr,Thomas-visc}. Despite
these discussions, no serious attempt to construct a gravity dual of
unitarity fermions has been made to date.


In this paper, we do not claim to have found the gravity dual of the
unitary Fermi gas.  However, we take the possible first step toward
such a duality.  We will construct a geometry whose symmetry coincides
with the Schr\"odinger symmetry~\cite{Hagen:1972pd,Niederer:1972},
which is the symmetry group of fermions at
unitarity~\cite{Mehen:1999nd}.  In doing so, we keep in mind that one
of the main evidences for gauge/gravity duality is the coincidence
between the conformal symmetry of the ${\cal N}=4$ field theory and
the symmetry of the AdS$_5$ space.  On the basis of this geometric
realization of the Schr\"odinger symmetry, we will be able to discuss
a nonrelativistic version of the AdS/CFT dictionary---the
operator-state correspondence, the relation between dimensions of
operators and masses of fields, etc.

The structure of this paper is as follows.  In
Sec.~\ref{sec:unitarity} we give a short introduction to fermions at
unitarity, emphasizing the field-theoretical aspects of the latter.
We also review the Schr\"odinger algebra.  In
Sec.~\ref{sec:Schroed} we describe how Schr\"odinger symmetry can be
embedded into a conformal symmetry in a higher dimension.  We consider
operator-field mapping in Sec.~\ref{sec:op-field}.  In
Sec.~\ref{sec:sources} we show how the conservation laws for mass,
energy and momentum are realized holographically.  We conclude with
Sec.~\ref{sec:conclusion}.

In this paper $d$ always refers to the number of spatial dimensions in
the nonrelativistic theory, so $d=3$ corresponds to the real world.

\section{Review of fermions at unitarity and Schr\"odinger symmetry}
\label{sec:unitarity}

In this section we collect various known facts about fermions at
unitarity and the Schr\"odinger symmetry.  The goal is not to present
an exhaustive treatment, but only to have a minimal amount of
materials needed for later discussions.  Further details can be found
in~\cite{Nishida:2007pj}.  We are mostly interested in vacuum
correlation functions (zero temperature and zero chemical potential), 
but not in the thermodynamics of the system at nonzero chemical potential.
The reasons are twofold: i) the chemical potential breaks the 
Schr\"odinger symmetry and ii) even at zero chemical potential there are
nontrivial questions, such as the spectrum of primary operators (see below).
We will comment on how chemical potential 
can be taken into account in Sec.~\ref{sec:conclusion}. 

One way to arrive at the theory of unitarity fermions is to start from
noninteracting fermions,
\begin{equation}
  {\cal L} = i\psi^\+\d_t \psi - \frac{|\nabla\psi|^2}{2m}\,,
\end{equation}
add a source $\phi$ coupled to the ``dimer'' field
$\psi_\downarrow\psi_\uparrow$~\cite{Nishida:2006br},
\begin{equation}\label{L-unit}
  {\cal L} = i\psi^\+\d_t \psi - \frac{|\nabla\psi|^2}{2m} 
  + \phi^*\psi_\downarrow\psi_\uparrow 
  + \phi \psi_\uparrow^\+ \psi_\downarrow^\+ ,
\end{equation}
and then promote the source $\phi$ to a dynamic field.  There is no
kinetic term for $\phi$ in the bare Lagrangian, but it will be
generated by a fermion loop.  Depending on the regularization scheme,
one may need to add to~(\ref{L-unit}) a counterterm
$c_0^{-1}\phi^*\phi$ to cancel the UV divergence in the one-loop
$\phi$ selfenergy (such a term is needed in momentum cutoff
regularization but not in dimensional regularization.)  The theory
defined by the Lagrangian~(\ref{L-unit}) is UV complete in spatial
dimension $2<d<4$, including the physically most relevant case of
$d=3$.  This system is called ``fermions at unitarity,'' which refers
to the fact that the $s$-wave scattering cross section between two
fermions saturates the unitarity bound.

Another description of fermions at unitarity is in terms of the
Lagrangian
\begin{equation}
  {\cal L}  = i\psi^\+\d_t \psi - \frac{|\nabla\psi|^2}{2m}
    - c_0 \psi^\+_\downarrow \psi^\+_\uparrow 
  \psi_\uparrow \psi_\downarrow.
\end{equation}
where $c_0$ is an interaction constant.
The interaction is irrelevant in spatial dimensions $d>2$, and is
marginal at $d=2$.  
At $d=2+\epsilon$
there is a nontrivial fixed point at a finite and negative value of
$c_0$ of order $\epsilon$~\cite{Sachdev}.  The situation is similar to
the nonlinear sigma model in $2+\epsilon$ dimensions.

In the quantum-mechanical language, unitarity fermions are defined as a
system with the free Hamiltonian
\begin{equation}
  H = \sum_i \frac{\mathbf{p}_i^2}{2m}\,,
\end{equation}
but with a nontrivial Hilbert space, defined to contain those
wavefunctions $\psi(\x_1,\x_2,\ldots;\y_1,\y_2,\ldots)$ (where $\x_i$
are coordinates of spin-up particles and $\y_j$ are those of spin-down
particles) which satisfy the following boundary conditions when a
spin-up and a spin-down particle approach each other,
\begin{equation}\label{unit-bc}
  \psi(\x_1,\x_2,\ldots;\y_1,\y_2,\ldots) \to 
  \frac{C}{|\x_i - \y_j|} + O(|\x_i-\y_j|).
\end{equation}
where $C$ depends only on coordinates other than $\x_i$ and $\y_j$.
This boundary condition can be achieved by letting the fermions
interact through some pairwise potential (say, a square-well
potential) that has one bound state at threshold.  In the limit of
zero range of the potential $r_0\to0$, keeping the zero-energy bound
state, the two-body wave function satisfies the boundary
condition~(\ref{unit-bc}) and the physics is universal.


Both free fermions and fermions at unitarity have the Schr\"odinger
symmetry---the symmetry group of the Schr\"odinger equation in free
space, which is the nonrelativistic version of conformal
symmetry~\cite{Mehen:1999nd}.  The generators of the Schr\"odinger
algebra include temporal translation $H$, spatial translations $P^i$,
rotations $M^{ij}$, Galilean boosts $K^i$, dilatation $D$ (where time
and space dilate with different factors: $t\to e^{2\lambda}t$, $\x\to
e^\lambda\x$), one special conformal transformation $C$ [which takes
$t\to t/(1+\lambda t)$, $\x\to \x/(1+\lambda t)$], and the mass
operator $M$.  The nonzero commutators are
\begin{equation}\label{Schroed-al}
\begin{split}
  & [M^{ij},\, M^{kl}] = i( \delta^{ik} M^{jl} + 
  \delta^{jl} M^{ik} - \delta^{il} M^{jk} - \delta^{jk} M^{il}),\\
  & [M^{ij},\, P^k] = i(\delta^{ik} P^j - \delta^{jk} P^i), \quad
  [M^{ij},\, K^k] = i(\delta^{ik} K^j - \delta^{jk} K^i), \\
  & [D,\, P^i] = -i P^i, \quad [D,\, K^i] = i K^i, \quad
    [P^i,\, K^j] = -i\delta^{ij} M,\\
  & [D,\, H] = -2iH, \quad [D,\, C] = 2iC,\quad
    [H,\, C] = iD.
\end{split}
\end{equation}
The theory of unitarity fermions is also symmetric under an SU(2)
group of spin rotations.

The theory of unitarity fermions is an example of nonrelativistic
conformal field theories (NRCFTs).  Many concepts of relativistic CFT,
such as scaling dimensions and primary operators, have counterparts in
nonrelativistic CFTs.  A local operator ${\cal O}$ is said to have
scaling dimension $\Delta$ if $[D,\, {\cal O}(0)]=-i\Delta{\cal
O}(0)$.  
Primary operators satisfy $[K^i,\, {\cal O}(0)]=[C,\,
{\cal O}(0)]=0$.  To solve the theory of unitarity fermions at zero
temperature and chemical potential is, in particular, to find the
spectrum of all primary operators.

In the theory of unitarity fermions, there is a quantum-mechanical
interpretation of the dimensions of primary 
operators~\cite{WernerCastin,Tan,Nishida:2007pj}.  A primary
operator with dimension $\Delta$ and charges $N_\uparrow$ and
$N_\downarrow$ with respect to the spin-up and spin-down particle
numbers (the total particle numbers is $N=N_\uparrow+N_\downarrow$)
corresponds to a solution of the zero-energy Schr\"odinger equation:
\begin{equation}
  \left(\sum_i\frac{\d^2}{\d x_i^2} + \sum_j\frac{\d^2}{\d y_j^2}\right)
  \psi(\x_1,\x_2,\ldots,\x_{N_\uparrow};\y_1,\y_2,\ldots,\y_{N_\downarrow}) 
  = 0,
\end{equation}
which satisfies the boundary condition~(\ref{unit-bc}) and with a
scaling behavior
\begin{equation}
  \psi(\x_1,\x_2,\ldots,\y_1,\y_2,\ldots) = R^\nu \psi(\Omega_k),
\end{equation}
where $R$ is an overall scale of the relative distances between
$\x_i$, $\y_j$, and $\Omega_k$ are dimensionless variables that are
defined through the ratios of the relative distances.  Equations~(7) and
(8) define, for given $N_\uparrow$ and $N_\downarrow$, a discrete set
of possible values for $\nu$.  For example, in three spatial
dimensions, for $N_\uparrow=N_\downarrow=1$, there are two possible
values for $\nu$: 0 and $-1$.  For $N_\uparrow=2$, $N_\downarrow=1$,
the lowest value for $\nu$ is $\approx-0.22728$.  Each value of $\nu$
corresponds to an operator with dimension $\Delta$, which is related
to $\nu$ by
\begin{equation}
  \Delta = \nu + \frac{dN}2\,.
\end{equation}
It has also been established that each primary operator corresponds to
a eigenstate of the Hamiltonian of unitarity fermion in an isotropic
harmonic potential of frequency
$\omega$~\cite{WernerCastin,Tan,Nishida:2007pj}.  The scaling
dimension of the operator simply coincides with the energy of the
state:
\begin{equation}
  E = \Delta\hbar\omega.
\end{equation}

The first nontrivial operator is the dimer
$\psi_\downarrow\psi_\uparrow$.  It has dimension $\Delta=d$ in the
free theory, and $\Delta=2$ in the theory of fermions at unitarity.
This corresponds to the fact that the lowest energy state of two
fermions with opposite spins in a harmonic potential is $d\hbar\omega$
in the case of free fermions and $2\hbar\omega$ for unitarity
fermions.

\section{Embedding the Schr\"odinger group into a conformal group}
\label{sec:Schroed}

To realize geometrically the Schr\"odinger symmetry, we first embed
the Schr\"odinger group in $d$ spatial dimensions Sch($d$) ($d=3$ for
the most interesting case of the unitarity Fermi gas) into the
relativistic conformal algebra in $d+2$ spacetime dimensions
O($d{+}2$, 2).
The next step will be to realize the Schr\"odinger
group as a symmetry of a $d+3$ dimensional spacetime background.  That
the Schr\"odinger algebra can be embedded into the relativistic
conformal algebra can be seen from the following.  Consider the
massless Klein-Gordon equation in $((d{+}1)+1)$-dimensional Minkowski
spacetime,
\begin{equation}
  \Box\phi \equiv -\d_t^2\phi + \sum_{i=1}^{d+1} \d_i^2\phi = 0.
\end{equation}
This equation is conformally invariant.  Defining the light-cone
coordinates,
\begin{equation}
  x^\pm = \frac{x^0 \pm x^{d+1}}{\sqrt2}\,,
\end{equation}
the Klein-Gordon equation becomes
\begin{equation}
  \left(-2\frac\d{\d x^-} \frac\d{\d x^+} 
  + \sum_{i=1}^{d}\d_i^2\right) \phi =0.
\end{equation}
If we make an identification $\d/\d x^-=-im$, then the
equation has the form of the Schr\"odinger equation in free space,
with the light-cone coordinate $x^+$ playing the role of time,
\begin{equation}
  \left( 2im \frac\d{\d x^+} + \d_i\d_i\right) \phi = 0.
\end{equation}
This equation has the Schr\"odinger symmetry Sch($d$).  Since the
original Klein-Gordon equation has conformal symmetry, this means
that Sch($d$) is a subgroup of O($d{+}2$, 2).

Let us now discuss the embedding explicitly.  The conformal algebra is
\begin{equation}\label{conform}
\begin{split}
  [\tilde M^{\mu\nu},\, \tilde M^{\alpha\beta}] 
      &= i(\eta^{\mu\alpha} \tilde M^{\nu\beta}
  + \eta^{\nu\beta} \tilde M^{\mu\alpha} - \eta^{\mu\beta} \tilde M^{\nu\alpha}
  - \eta^{\nu\alpha} \tilde M^{\mu\beta} ),\\
  [\tilde M^{\mu\nu},\, \tilde P^\alpha] &= i (\eta^{\mu\alpha}\tilde P^\nu -
   \eta^{\nu\alpha} \tilde P^\mu),\\
  [\tilde D,\, \tilde P^\mu] &= -i \tilde P^\mu, \quad 
     [\tilde D,\, \tilde K^\mu] = i \tilde K^\mu, \\
  [\tilde P^\mu,\, \tilde K^\nu] 
     &= -2i(\eta^{\mu\nu} \tilde D + \tilde M^{\mu\nu}),
\end{split}
\end{equation}
where Greek indices run $0,\ldots,d+1$, and all other commutators are
equal to 0.  The tilde signs denote relativistic operators; we reserve
untilded symbols for the nonrelativistic generators.  We identify the
light-cone momentum $\tilde P^+=(\tilde P^0+\tilde P^{d+1})/\sqrt2$
with the mass operator $M$ in the nonrelativistic theory.  We
now select all operators in the conformal algebra that commute with
$\tilde P^+$.  Clearly these operators form a closed algebra, and it is easy
to check that it is the Schr\"odinger algebra in $d$ spatial
dimensions.  The identification is as follows:
\begin{equation}\label{embedding}
\begin{split}
  & M = \tilde P^+,\quad H = \tilde P^-, \quad P^i = \tilde P^i,\quad
    M^{ij} = \tilde M^{ij}, \\ 
  & K^i = \tilde M^{i+}, \quad D = \tilde D + \tilde M^{+-}, \quad
  C = \frac{\tilde K^+}2\,.
\end{split}
\end{equation}
From Eqs.~(\ref{conform}) and (\ref{embedding}) one finds the
commutators between the untilded operators to be
exactly the Schr\"odinger algebra, Eqs.~(\ref{Schroed-al}).

\section{Geometric realization of the Schr\"odinger symmetry}
\label{sec:geom-rel}

To realize the Schr\"odinger symmetry geometrically, we will take the
AdS metric, which is is invariant under the whole conformal group, and
then deform it to reduce the symmetry down to the Schr\"odinger group.
The AdS space, in Poincar\'e coordinates, is
\begin{equation}
  ds^2 = \frac{\eta_{\mu\nu}dx^\mu dx^\nu + dz^2}{z^2}\,.
\end{equation}
The generators of the conformal group correspond to the following
infinitesimal coordinates transformations that leave the metric
unchanged,
\begin{equation}\label{conf-act}
\begin{split}
  P^\mu: &~ x^\mu \to x^\mu + a^\mu, \\
  D: &~ x^\mu \to (1-a) x^\mu, \quad z \to (1-a)z,\\
  K^\mu : &~ x^\mu \to x^\mu + a^\mu (z^2 + x \cdot x)
          - 2x^\mu (a\cdot x)
\end{split}
\end{equation}
(here $x\cdot x\equiv\eta_{\mu\nu}x^\mu x^\nu$).  

We will now deform the metric so to reduce the symmetry to the
Schr\"odinger group.  In particular, we want the metric to be
invariant under $D=\tilde D+\tilde M^{+-}$, which is a linear
combination of a boost along the $x^{d+1}$ direction $\tilde M^{+-}$
and the scale transformation $\tilde D$, but not separately under
$\tilde M^{+-}$ or $\tilde D$.  The following metric satisfies this
condition:
\begin{equation}\label{schroed-met}
  ds^2 =  -\frac{2(dx^+)^2}{z^4} + \frac{-2dx^+ dx^- + dx^i dx^i + dz^2}{z^2}
  \,.
\end{equation}
It is straightforward to verify that the metric~(\ref{schroed-met})
exhibits a full Schr\"odinger symmetry.  From Eqs.~(\ref{embedding})
and (\ref{conf-act}) one finds that the generators of the
Schr\"odinger algebra correspond to the following isometries of the
metric:
\begin{equation}
\begin{split}
  P^i :& ~ x^i \to x^i + a^i,\quad 
  H: ~ x^+ \to x^+ + a,  \quad M: ~ x^- \to x^- + a, \\
  K^i :& ~   x^i \to x^i - a^i x^+, \quad x^- \to x^- - a^i x^i,\\
  D :& ~ x^i \to (1-a)x^i, \quad z\to (1-a)z, \quad
         x^+ \to (1-a)^2 x^+, \quad x^- \to x^-,\\
  C: & ~ z \to (1 - ax^+) z, \quad x^i \to (1 - ax^+) x^i, \quad
      x^+ \to (1 - ax^+) x^+, \\\
     &\qquad   x^- \to x^- - \frac a2 (x^i x^i + z^2).
\end{split}
\end{equation}

We thus hypothesize that the gravity dual of the unitarity Fermi gas
is a theory living on the background metric~(\ref{schroed-met}).
Currently we have very little idea of what this theory is.  We shall
now discuss several issues related to this proposal.

\medskip

i) The mass $M$ in the Schr\"odinger algebra is mapped onto the
light-cone momentum $P^+\sim\d/\d x^-$.  In nonrelativistic theories
the mass spectrum is normally discrete: for example, in the case of
fermions at unitarity the mass of any operator is a multiple of the
mass of the elementary fermion.  It is possible that the light-cone
coordinate $x^-$ is compactified, which would naturally give rise to
the discreteness of the mass spectrum.

\medskip

ii) In AdS/CFT correspondence the number of color $N_c$ of the field
theory controls the magnitude of quantum effects in the string theory
side: in the large $N_c$ limit the string theory side becomes a
classical theory.  The usual unitarity Fermi gas does not have this
large parameter $N$, hence the dual theory probably has unsuppressed
quantum effects.  However, there exists an extension of the unitarity
Fermi gas with Sp($2N$) symmetry~\cite{Sachdev,Radzihovsky}.  The
gravity dual of this theory may be a classical theory in the limit of
large $N$, although with an infinite number of fields, similar to
the conjectured dual of the critical O($N$) vector model in 2+1
dimensions~\cite{Klebanov:2002ja}.

\medskip

iii) We can write down a toy model in which the
metric~(\ref{schroed-met}) is a solution to field equations.  Consider
the theory of gravity coupled to a massive vector field with a
negative cosmological constant,
\begin{equation}\label{model}
  S = \int\!d^{d+2}x\,dz\, \sqrt{-g} \left(\frac12 R - \Lambda -
      \frac14 H_{\mu\nu} H^{\mu\nu} - \frac{m^2}2 C_\mu C^\mu \right),
\end{equation}
where $H_{\mu\nu}=\d_\mu C_\nu-\d_\nu C_\mu$.
One can check that Eq.~(\ref{schroed-met}), together with
\begin{equation}\label{A-bg}
  C^- = 1,
\end{equation}
is a solution to the coupled Einstein and Proca equations for the
following choice of $\Lambda$ and $m^2$:
\begin{equation}\label{A-mass}
  \Lambda=-\frac12(d+1)(d+2),\qquad m^2=2(d+2).
\end{equation}

\medskip

iv) Although the $g_{++}$ metric component has $z^{-4}$ singularity at
$z=0$, the metric has a plane-wave form and all scalar curvatures are
finite.  For example, the most singular component of the Ricci tensor,
$R_{++}$, has a $z^{-4}$ singularity, as the $C_{+i+i}$ and $C_{+z+z}$
components of the Weyl tensor.
However, since $g^{++}=0$, any scalar constructed from the curvature tensor
is regular.  

\medskip

v) In terms of a dual field theory, the field $A_\mu$ with mass in
Eq.~(\ref{A-mass}) corresponds to a vector operator $O^\mu$ with
dimension $\Delta$, which can be found from the general formula
\begin{equation}
  (\Delta-1)[\Delta+1-(d+2)] = 2(d+2),
\end{equation}
from which $\Delta=d+3$.  We thus can think about the quantum field
theory as an irrelevant deformation of the original CFT, with the
action
\begin{equation}
  S = S_{\rm CFT} + J\!\int\!d^{d+2}x\, O^+ .
\end{equation}

\section{Operator-field correspondence}
\label{sec:op-field}

Let us now discuss the relationship between the dimension of operators
and masses of fields in this putative nonrelativistic AdS/CFT
correspondence.  Consider an operator $O$ dual to a massive scalar
field $\phi$ with mass $m_0$.  We shall assume that it couples minimally
to gravity,
\begin{equation}
  S = -\int\! d^{d+3}x\, \sqrt{-g}
  (g^{\mu\nu} \d_\mu\phi^* \d_\nu\phi + m_0^2 \phi^*\phi).
\end{equation}
Assuming the light-cone coordinate $x^-$ is periodic, let us concentrate
only on the Kaluza-Klein mode with $P^+=M$.  The action now
becomes
\begin{equation}
  S = \int\!d^{d+2}x\, dz\, \frac1{z^{d+3}} \left(
  2iM z^2 \phi^*\d_t\phi - z^2\d_i\phi^*\d_i \phi - m^2\phi^*\phi\right),
\end{equation}
where the ``nonrelativistic bulk mass'' $m^2$ is related to the
original mass $m_0^2$ by $m^2=m_0^2+2M^2$.  Contributions to $m^2$ can
arise from interaction terms between $C_\mu$ and $\phi$, for example
$|C^\mu\d_\mu\phi|^2$, $|C^\mu C^\nu\d_\mu\d_\nu\phi|^2$, etc.  We
therefore will treat $m^2$ as an independent parameter.

The field equation for $\phi$ is
\begin{equation}
  \d_z^2 \phi - \frac{d{+}1}z\d_z \phi + \left( 2M\omega - \vec k^2
  - \frac{m^2}{z^2}\right)\phi = 0.
\end{equation}
The two independent solutions are
\begin{equation}
  \phi_\pm = z^{d/2+1} K_{\pm\nu} (pz), \qquad p = (\vec k^2 - 2M\omega)^{1/2},
  \qquad \nu =  \sqrt{m^2 + \frac{(d{+}2)^2}4}\,.
\end{equation}

As in usual AdS/CFT correspondence, one choice of $\phi_\pm$
corresponds to turning a source for $O$ in the boundary theory,
and another choice corresponds to a condensate of $O$.  One can
distinguish two cases:
\begin{enumerate}
\item When $\nu\geq1$, $\phi_+$ is non-normalizable and $\phi_-$ is
renormalizable.  Therefore $\phi_+$ corresponds to the source and
$\phi_-$ to the condensate.  The correlation function of $O$ is
\begin{equation}
  \< OO\> \sim (\vec k^2 - 2M\omega)^{2\nu},
\end{equation}
which translate into the scaling dimension
\begin{equation}
  \Delta = \frac {d{+}2}2 + \nu .
\end{equation}

\item When $0<\nu<1$ both asymptotics are normalizable, and there is
an ambiguity in the choice of the source and condensate boundary
conditions.  These two choices should correspond to two different
nonrelativistic CFTs.  In one choice the operator $O$ has dimension
$\Delta=(d+2)/2+\nu$, and in the other choice $\Delta=(d+2)/2-\nu$.
It is similar to the situation discussed in~\cite{Klebanov:1999tb}.
\end{enumerate}

The smallest dimension of an operator one can get is
$\Delta=(d+2)/2-\nu$ when $\nu\to1$.  Therefore, there is a lower
bound on operator dimensions,
\begin{equation}\label{unitary}
  \Delta > \frac d2 \,.
\end{equation}
This bound is very natural if one remember that operator dimensions
correspond to eigenvalues of the Hamiltonian in an external harmonic
potential.  For a system of particles in a harmonic potential, one can
separate the center-of-mass motion from the relative motion.
Equation~(\ref{unitary}) means that the total energy should be larger than
the zero-point energy of the center-of-mass motion.

The fact that there are pairs of nonrelativistic conformal field
theories with two different values of the dimensions of $O$ is a
welcome feature of the construction.  In fact, free fermions and
fermions at unitarity can be considered as such a pair.  In the theory
with free fermions the operator $\psi_\downarrow\psi_\uparrow$ has
dimension $d$, and for unitarity fermions, this operator has dimension
2.  The two numbers are symmetric with respect to $(d+2)/2$:
\begin{equation}
  d = \frac{d+2}2 + \frac{d-2}2, \qquad 2 = \frac{d+2}2 - \frac{d-2}2\,.
\end{equation}
Therefore, free fermions and fermions at unitarity should correspond
to the same theory, but with different interpretations for the
asymptotics of the field dual to the operator
$\psi_\downarrow\psi_\uparrow$.

A similar situation exists in the case of Fermi gas at unitarity with
two different masses for spin-up and spin-down
fermions~\cite{Nishida:2007mr}.  In a certain interval of the mass
ratios (between approximately 8.6 and 13.6), there exist two different
scale-invariant theories which differ from each other, in our
language, by the dimension of a three-body $p$-wave operator.  At the
upper end of the interval (mass ratio 13.6) the dimension of this
operator tends to 5/2 in both theories; at the lower end it has
dimension $3/2$ in the theory with three-body resonance and $7/2$ in
the theory without three-body resonance.

\section{Turning on sources}
\label{sec:sources}

Let us now try to turn on sources coupled to conserved currents in the
boundary theory.  That would correspond to turning on non-normalizable
modes.  For the fields that enter the model action~(\ref{model}), the
general behavior of the non-normalizable part of the metric and the
field $C_\mu$ near $z=0$ is
\begin{equation}\label{bc}
\begin{split}
   ds^2 &= -\frac{2e^{-2\Phi}}{z^4}(dx^+{-}B_i dx^i)^2
  - \frac{2e^{-\Phi}}{z^2}(dx^+ {-} B_i dx^i)(dx^- {-} A_0 dx^+ {-} A_i dx^i)\\
  &\qquad\qquad + \frac{g_{ij} dx^i dx^j + dz^2}{z^2} + O(z^0),\\ 
   C^- &= 1.
\end{split}
\end{equation}
We have chosen the gauge $g_{\mu z}=0$.  The non-normalizable metric
fluctuations are parametrized by the functions $A_0$, $A_i$, $\Phi$,
and $B_i$ of $x^+\equiv t$ and $x^i$.  These functions are interpreted
as background fields, on which the boundary theory exists.
Following the general philosophy of AdS/CFT correspondence, we assume
that the partition function of the high-dimensional theory with the
boundary condition~(\ref{bc}) is equal to the partition function of an
NRCFT in the background fields,
\begin{equation}
  Z = Z[A_0, A_i, \Phi, B_i, g_{ij}].
\end{equation}
This partition function should be invariant with respect to a group of
gauge transformations acting on the background fields, which we will
derive.

The gauge condition $g_{\mu z}=0$ does not completely fix the metric:
there is a residual gauge symmetry parametrized by arbitrary functions
of $t$ and $x^i$ (but not of $z$):
\begin{equation}\label{residual}
  t\to t'= t+ \xi^t(t,\x), \quad x^- \to x^{-\prime}=x^- + \xi^-(t,\x), \quad
  x^i \to x^{i\prime}=x^i + \xi^i(t,\x),
\end{equation}
and another set of infinitesimal transformations characterized by a
function $\omega(t,\x)$,
\begin{equation}\label{residual2}
  z \to z' = z - \omega(t,\x) z, \qquad
  x^\mu \to x^{\mu\prime} = x^\mu + \frac12 g^{\mu\nu}\d_\nu\omega.
\end{equation}

Consider first~(\ref{residual}).  Under these residual gauge
transformations, the fields entering the metric~(\ref{bc}) change in
the following way:
\begin{equation}\label{new-gci}
\begin{split}
  \delta A_0& = \dot\xi^- - A_0\dot\xi^t - A_i\dot\xi^i -\xi^\mu\d_\mu A_0, \\
  \delta A_i &= \d_i\xi^- - A_0\d_i\xi^t - e^\Phi g_{ij}\dot\xi^j
     -\xi^\mu\d_\mu A_i - A_j \d_i \xi^j,\\
  \delta\Phi &=  \dot \xi^t - B_i\dot \xi^i-\xi^\mu\d_\mu\Phi,\\
  \delta B_i &= \d_i\xi^t + B_i(\dot\xi^t - B_j\dot\xi^j) 
      -\xi^\mu\d_\mu B_i - B_j \d_i \xi^j,\\
  \delta g_{ij} &= -(B_ig_{jk}+B_jg_{ik})\dot\xi^k -\xi^\mu\d_\mu g_{ij}
    -g_{kj}\d_i\xi^k - g_{ik}\d_j\xi^k,
\end{split}
\end{equation}
where $\xi^\mu\d_\mu\equiv\xi^t\d_t+\xi^i\d_i$.  The residual gauge
symmetry implies that the partition function of the boundary theory
should be invariant under such transformations,
\begin{equation}
  \delta Z = 0.
\end{equation}

Can one formulate NRCFTs on background fields with this symmetry?  In
fact, it can be done explicitly in the theory of free nonrelativistic
particles.  One introduces the interaction to the background fields in
the following manner:
\begin{multline}\label{SB}
  S = \int\!dt\,d\x\, \sqrt{g}\, e^{-\Phi} \left[ \frac i2 e^\Phi
    (\psi^\+ D_t\psi - D_t\psi^\+\psi) -\frac{g^{ij}}{2m} D_i\psi^\+
    D_j \psi \right. \\ \left.  - \frac{B^i}{2m}(D_t\psi^\+ D_i\psi
    +D_i\psi^\+ D_t\psi) -\frac{B^2}{2m} D_t\psi^\+ D_t\psi \right],
\end{multline}
where $g^{ij}$ is the inverse matrix of $g_{ij}$, $g\equiv\det |g_{ij}|$,
$B^i\equiv g^{ij}B_i$, $B^2\equiv B^iB_i$, 
and $D_\mu\psi\equiv\d_\mu\psi-imA_\mu\psi$.  One can verify
directly that the action~(\ref{SB}) is invariant under the
transformations~(\ref{new-gci}), if $\psi$ transforms as
\begin{equation}
  \delta\psi = im\xi^-\psi - \xi^\mu \d_\mu\psi.
\end{equation}
In fact, this invariance is an extension of the general coordinate
invariance previously discussed in~\cite{Son:2005rv}.  The invariance
found in~\cite{Son:2005rv} corresponds to restricting
$\Phi=B_i=\xi^t=0$ in all formulas.

To linear order in external field, the action is
\begin{equation}
  S = S[0]+ \int\!dt\,d\x\, \left( A_0\rho + A_i j^i + \Phi\epsilon
      + B_i j^i_\epsilon + \frac12 h_{ij}\Pi^{ij}\right),
\end{equation}
and from Eq.~(\ref{SB}) one reads out the physical meaning of the
operators coupled to the external sources:
\begin{itemize}
\item $h_{ij}$ is coupled to the stress tensor $\Pi^{ij}$,
\item $A_\mu$ is coupled to the mass current $(\rho,\, \mathbf{j})$,
\item $(\Phi,\,B_i)$ are coupled to the energy current 
$(\epsilon,\, {\bf j}_\epsilon)$.
\end{itemize}
The invariance of the partition function with respect to the gauge
transformations~(\ref{new-gci}) leads to an infinite set of
Takahashi-Ward identities for the correlation functions.  The simplest
ones are for the one-point functions.  The fact that the group of
invariance includes gauge transformation of $A_\mu$: $\delta
A_\mu=\d_\mu\xi^-$ guarantees the conservation of mass.  The fact that
the linear parts in the transformation laws for $\Phi$ and $B_i$ look
like a gauge transformation, $\delta\Phi=\dot\xi^t+\cdots$ and $\delta
B_i=\d_i\xi^t+\cdots$ leads to energy conservation in the absence of
external fields:
\begin{equation}
  \left.
  \d_t\left\langle \frac{\d\ln Z}{\d\Phi}\right\rangle +
  \d_i\left\langle \frac{\d\ln Z}{\d B_i}\right\rangle
   \right|_{A_\mu=\Phi=B_i=h_{ij}=0} = 0.
\end{equation}
Energy is not conserved in a general background (which is natural,
since the background fields exert external forces on the system).
Similarly, momentum conservation $\d_t j^i+\d_j\Pi^{ij}=0$ (and the
fact that momentum density coincides with mass current) is related to
terms linear in $\xi^i$ in $\delta A_i$ and $\delta g_{ij}$: $\delta
A_i=-\dot\xi^i+\cdots$, $\delta g_{ij}=-\d_i\xi^j -
\d_j \xi^i +\cdots$.

Let us now turn to the transformations~(\ref{residual2}),
under which
\begin{equation}\label{deltaPhig}
  \delta\Phi= 2\omega, \qquad \delta g_{ij} = -2\omega g_{ij}.
\end{equation}
The invariance of the partition function with respect to this
transformation implies
\begin{equation}
  2\epsilon = \Pi^i_i,
\end{equation}
which is the familiar relationship between energy and pressure, 
\begin{equation}
  E = \frac d2 PV,
\end{equation}
valid for free gas as well as for Fermi gas at unitarity.  The
action~(\ref{SB}) is not invariant under~(\ref{deltaPhig}), but it can be
made so by replacing the ``minimal coupling'' by a ``conformal
coupling'' to external fields.  Therefore, the proposed holography 
is consistent
with conservation laws and the universal thermodynamic relation
between energy and pressure.

\section{Conclusion}
\label{sec:conclusion}

The main goal of the paper is to construct a geometry with the
symmetry of the Schr\"odinger group.  The existence of such a
geometrical realization make it possible to discuss the possibility of
a dual description of Fermi gas at unitarity at a concrete level.  It
remains to be seen if holography is a notion as useful in
nonrelativistic physics as it is for relativistic quantum field
theories.  At the very least, one should expect holography to provide
toy models with Schr\"odinger symmetry.

In this paper we have considered only the properties of the vacuum correlation
functions.  In order to construct the gravity dual of the
finite-density ground state, about which a lot is known both 
experimentally and theoretically,
one should turns on a background $A_0$ in the
metric~(\ref{bc}).  Superfluidity of the system should be encoded in the
condensation of the scalar field $\psi_\uparrow\psi_\downarrow$ (whose
dimension is 2 in the case of unitarity fermions, 
cf.~\cite{Hartnoll:2008vx,Gubser:2008zu}).
It would be interesting to find black-hole metrics
which realize nonrelativistic hydrodynamics and superfluid
hydrodynamics.  We defer this problem to future work.

\acknowledgments

The author thanks A.~Karch and Y.~Nishida for discussions leading to
this work, and S.~Hartnoll, V.~Hubeny, D.~Mateos, H.~Liu,
K.~Rajagopal, M.~Rangamani, S.~Shenker, and M.~Stephanov
for valuable comments.  This work is supported, in
part, by DOE Grant DE-FG02-00ER41132.

\medskip

{\sl Note added}---After this work was completed, J.~McGreevy informed
the author that he and K.~Balasubramanian have also obtained the
metric~(\ref{schroed-met}) and determined that it has nonrelativistic
conformal symmetry~\cite{McGreevy}.

\end{document}